\documentclass[12pt,preprint]{aastex}

\usepackage{amsmath}

\begin{document}


\title{MOA-2010-BLG-328Lb: a sub-Neptune orbiting very late M dwarf?}

\author{
K.~Furusawa\altaffilmark{1,A},
A.~Udalski\altaffilmark{2,B},
T.~Sumi\altaffilmark{3,1,A},
D.P.~Bennett\altaffilmark{4,A,C},
I.A.~Bond\altaffilmark{5,A},
A.~Gould\altaffilmark{6,D},
U.G.~J{\o}rgensen\altaffilmark{7,8,E},
C.~Snodgrass\altaffilmark{9,10,F,E}, 
D.~Dominis Prester\altaffilmark{11,C},
M.D.~Albrow\altaffilmark{12,C}
% ...
\\ (The PLANET Collaboration) \\
}

\altaffiltext{1}{Solar-Terrestrial Environment Laboratory, Nagoya University, Nagoya, 464-8601, Japan; furusawa@stelab.nagoya-u.ac.jp}
\altaffiltext{2}{Warsaw University Observatory, Al. Ujazdowskie 4, 00-478 Warszawa, Poland}
\altaffiltext{3}{Department of Earth and Space Science, Graduate School of Science, Osaka University, 1-1 Machikaneyama-cho, Toyonaka, Osaka 560-0043, Japan}
\altaffiltext{4}{Department of Physics, 225 Nieuwland Science Hall, University of Notre Dame, Notre Dame, IN 46556, USA}
\altaffiltext{5}{Institute for Information and Mathematical Sciences, Massey University, Private Bag 102-904, Auckland 1330, New Zealand}
\altaffiltext{6}{Department of Astronomy, Ohio State University, 140 West 18th Avenue, Columbus, OH 43210, USA}
\altaffiltext{7}{Niels Bohr Institutet, K{\o}benhavns Universitet, Juliane Maries Vej 30, 2100 Copenhagen, Denmark}
\altaffiltext{8}{Centre for Star and Planet Formation, Geological Museum, {\O}ster Voldgade 5, 1350 Copenhagen, Denmark}
\altaffiltext{9}{Max Planck Institute for Solar System Research, Max-Planck-Str. 2, 37191 Katlenburg-Lindau, Germany}
\altaffiltext{10}{European Southern Observatory, Casilla 19001, Vitacura 19, Santiago, Chile}
\altaffiltext{11}{Department of Physics, University of Rijeka, Omladinska 14, 51000 Rijeka, Croatia}
\altaffiltext{12}{University of Canterbury, Department of Physics and Astronomy, Private Bag 4800, Christchurch 8020, New Zealand}
% ...
\altaffiltext{A}{Microlensing Observations in Astrophysics (MOA) Collaboration}
\altaffiltext{B}{Optical Gravitational Lensing Experiment (OGLE)}
\altaffiltext{C}{Probing Lensing Anomalies Network (PLANET) Collaboration}
\altaffiltext{D}{Microlensing Follow Up Network ($\mu$FUN) Collaboration}
\altaffiltext{E}{Microlensing Network for Detection of Small Terrestrial Exoplanets (MiNDSTEp) Consortium}
\altaffiltext{F}{RoboNet Collaboration}

\begin{abstract}
We analyze the planetary microlensing event MOA-2010-BLG-328.  The
best fit yields host and planetary masses of
$M_{\rm{h}} = 0.11\pm 0.01\,M_\odot$ and $M_{\rm{p}}=9.2\pm 2.2\,M_\oplus$, 
corresponding
to a very late M dwarf and sub-Neptune-mass planet, respectively.  The
system lies at $D_{\rm{L}} = 0.81\pm 0.10\,$kpc with projected separation
$r_\perp=0.92\pm 0.16\,$AU.  Because of the host's a-priori-unlikely close
distance, as well as the unusual nature of the system, we consider
the possibility that the microlens parallax signal, which determines
the host mass and distance, is actually due to xallarap (source
orbital motion) that is being misinterpreted as parallax.
We show a result that favors the parallax solution,
even given its close host distance.
We show that future high-resolution astrometric measurements could decisively resolve the
remaining ambiguity of these solutions.
\end{abstract}

\keywords{gravitational lensing: micro - planetary systems}

\section{Introduction}
To date, more than 800 exoplanets have been discovered via several different methods.
Most of the exoplanets have been discovered with the
radial velocity \citep{Lovis2011} and transit methods \citep{Winn2011}.
These methods are most sensitive to planets in very close orbits, and
as a result, our understanding of the properties of exoplanetary 
systems is dominated by planets in close orbits.
While the number of exoplanet discoveries by microlensing is relatively small (18 discoveries
to date) \citep{Bond2004, Bachelet2012}, microlensing is sensitive to planets beyond the 
``snow-line" at $\sim 2.7\,{\rm AU}(M/M_{\sun})$ \citep{Kennedy2008}, where $M$ is the mass of the host star.
This region beyond the ``snow-line" is thought to be the dominant 
exoplanet birthplace, and microlensing is able to find planets down
to an Earth-mass \citep{Bennett1996} in this region.
Microlensing does not depend on the detection of any light from the
exoplanet host stars, so planets orbiting faint hosts, like
brown dwarfs and M-dwarfs \citep{Udalski2005,Dong2009,bennett08,moa192_naco,Batista2011}, can be detected.

Microlensing is one of several methods that has contributed to our
statistical understanding of the exoplanet distribution.
In other methods, \citet{Cumming2008} analyzed on 8 years of radial velocity measurements to
constraint the frequency of the Jupiter-mass planets (0.3 - 10 $M_{\rm{Jupiter }}$) 
with orbital periods of less than 2000 days, and found that less than 10.5\% of
the stars in their sample had such planets. \citet{Wittenmyer2011} 
used a 12-year radial velocity sample to search for Jupiter analogs,
and found that between 3.3\% and 37.2 \% of the stars in their
sample had such a planet with a planet.
When it comes to the transit method, \citet{Howard2012} reported the distribution of planets
as a function of planet radius, orbital period, and stellar effective temperature
for orbital periods less than 50 days around solar-type stars.
They measured an occurrence of $0.165 \pm 0.008$ planets per star for planets with radii $2-32 R_{\earth}$.
Microlensing has already demonstrated the ability to find both
Jupiter and Saturn-analog planets with the discovery of the
Jupiter/Saturn analog system, OGLE-2006-BLG-109Lb,c 
\citep{Gaudi2008, Bennett2010}. There have been several recent
papers that have looked at the statistical implications of the
microlensing exoplanet discoveries.
\citet{Sumi2010} determined the slope of the exoplanet mass
function beyond the snow line, and found that 
the cold Neptunes are $\sim 7$ times more common
than Jupiters. (The 95\% confidence level limit is more than three times more common.)
\citet{Gould2010} used six microlensing discoveries to show that low-mass
gas giant planets are quite common beyond the snow line of low-mass stars at
a level that is consistent with an extrapolation of the \citet{Cumming2008}
radial velocity results.
Most recently, \citet{Cassan2012} estimated the fraction of bound planets at 
separations of 0.5 - 10 AU with a somewhat larger microlensing sample.
They found that $17^{+6}_{-9}$ \% of stars host Jupiter-mass planets, while
$52^{+22}_{-29}$ \% and $62^{+35}_{-37}$ \% of stars host 
Neptune-mass planets (10 - 30 $M_{\oplus}$) and super-Earths (5 - 10 $M_{\oplus}$), 
respectively.

In this paper, we present the analysis of 
the planetary microlensing event MOA-2010-BLG-328. 
Section \ref{sec:observations} describes the observation of the event, and
the light curve modeling is presented in Section \ref{sec:modeling}.
The light curve shows evidence of orbital motion of either the Earth,
known as microlensing parallax, or of the source stars, which is often
referred to as the xallarap effect. These two possibilities have very
different implications for the properties of the host star and its planet.
However, we need the angular radius of the source star to work out the
implications for the star plus planet lens system, so we determine the
source star angular radius in Section \ref{sec:source}.
In Section \ref{sec:lens}, we present the implications for the properties
of the host star and its planet for both the parallax and xallarap solutions.
While the data do prefer the parallax solution,
the xallarap model is not completely excluded.
In section \ref{sec:discussion}, we describe how future
follow-up observations can distinguish between these two solutions,
and give our conclusions. 

\section{Observations}
\label{sec:observations}

Several groups search for exoplanets using the microlensing method, using
two different observing modes: wide field-of-view (FOV) surveys, and
narrow FOV follow-up observations.
The microlensing surveys active in 2010 were the Microlensing Observations in 
Astrophysics \citep[MOA;][]{Bond2001,Sumi2003} group and the Optical Gravitational Lensing 
Experiment \citep[OGLE;][]{Udalski2003}.
MOA uses the 1.8m MOA-II telescope equipped with the
10k $\times$ 8k-pixel CCD camera MOA-cam3 \citep{Sako2008} with 
a 2.2 deg$^{2}$ FOV to monitor $\sim$ 44 deg$^{2}$ of the Galactic bulge
with a cadence of one observation of each field every 15-95 minutes depending on the field.
MOA identifies microlensing events in real time \citep{Bond2001}, and announced
607 microlensing alerts in 2010. In 2010, the OGLE collaboration initiated
the OGLE-IV survey, after upgrading their CCD camera from the OGLE-III
system which operated from 2001-2009.
OGLE observations are conducted with 1.3m Warsaw telescope at Las Campanas Observatory, Chile.
The OGLE-IV survey employs a 1.4 deg$^{2}$, 256 Mega pixel, 32 chip CCD mosaic camera to survey an even larger area of
the Galactic bulge at cadences ranging from one observation every 20 minutes to less than
one observation per day.
The OGLE real-time event detection system,
known as the Early Warning System (EWS) \citep{Udalski2003}, was not operational
in 2010.

The follow-up groups employ narrow FOV telescopes spread across the world (mostly in the Southern Hemisphere) for
high cadence photometric monitoring of a subset of microlensing events found by the survey groups.
Generally, the events observed by the follow-up groups are events
with a high planet detection sensitivity \citep{griest98, Horne2009} or events
in which a candidate planetary signal has been seen.
Follow-up groups include 
the Microlensing Follow-Up Network \citep[$\mu$FUN;][]{Gould2006}, 
Microlensing Network for the Detection of Small Terrestrial Exoplanets \citep[MiNDSTEp;][]{Dominik2010}, 
RoboNet \citep{Tsapras2009},
and 
the Probing Lensing Anomalies NETwork \citep[PLANET;][]{Beaulieu2006}.
Because the planetary deviations are short, with durations ranging from
a few hours to a few days, high cadence observations from observatories
widely spaced in longitude are needed to provide good sampling.

The microlensing event MOA-2010-BLG-328 
(R.A., decl.)(J2000) = (17$^{h}$57$^{m}$59$^{s}$.12, -30$^{\circ}$42'54".63)[($l, b$) = 
(-0$^{\circ}$.16, -3$^{\circ}$.21)]
was detected 
and alerted by MOA on 16 June 2010 (HJD' $\equiv$ HJD - 2450000 $\sim$ 5363). 
The MOA observer noticed a few data points at HJD' $\sim$ 5402
that were above the prediction of the single lens light curve model,
but waited for the next observations, three days later, for the significance
of the deviation to reach the threshold to issue an anomaly alert.
This anomaly alert was issued to the other microlensing groups 
at UT 11:30 27 July (HJD' $\sim$ 5405).
One day later, MOA circulated a preliminary planetary model, and shortly thereafter,
observations were begun by the follow-up groups.
Follow-up data was obtained from the $\mu$FUN, PLANET,
MiNDSTEp, and RoboNet groups. 
$\mu$FUN obtained data from the CTIO 1.3 m telescope 
in Chile in the $I$, $V$, and $H$-bands, the Palomar Observatory 1.5 m
telescope in USA in the $I$-band, and the 
Farm Cove Observatory 0.36 m in New Zealand in unfiltered pass band.
$\mu$FUN also obtained data from Auckland Observatory, Kumeu Observatory,
and Possum Observatory, all in New Zealand; unfortunately, they obtained
only one night of observations, so these data are not used in the analysis.
The data sets from PLANET consist of $V$ and $I$-band data
from SAAO 1.0m telescope in South Africa,
and $I$-band data from the Canopus Observatory 1.0 m telescope in Australia.
Robonet provided data from
the Faulkes Telescope North 2.0 m in Hawaii in the $V$ and $I$-bands,
the Faulkes Telescope South 2.0 m in Australia in the $I$-band, and
the Liverpool Telescope 2.0 m in Canary Islands in the $I$-band.
The MiNDSTEp group provided data from the 
Danish 1.54 m telescope at ESO La Silla in Chile in the $I$-band. 
MOA's observations were done in the wide MOA-red band, which is
approximately equivalent to $R+I$, and the observations during the
main peak of the planetary deviation
were taken at a cadence of about one image every 10 minutes, or 4-5 times
higher than the normal observing cadence due to a high magnification
event in the same field and the detection of the anomaly in this event.
Due to poor weather at the Mt. John University Observatory,
where the MOA telescope is located,
the planetary signal was recognized and
announced after it was already nearing the second peak, so the
light curve coverage from the follow-up groups is poor. 
Fortunately, much of the early part of the planetary deviation was monitored
by the OGLE-IV survey (in the $I$-band), and so we have good coverage of most of 
the planetary deviation from the MOA and OGLE surveys.

Most of the photometry was done by the standard difference imaging photometry 
method for each group.  The MOA data were reduced with the MOA Difference 
Image Analysis (DIA) pipeline \citep{Bond2001}, and the OGLE data were 
reduced with the OGLE DIA photometry pipeline \citep{Udalski2003}.
The photometry of $\mu$FUN and PLANET was performed with the
PLANET group's PYSYS \citep{Albrow2009} difference imaging code.
The RoboNet data were reduced with DanDIA \citep{Bramich2008}, and the
MiNDSTEp data were reduced with DIAPL \citep{Wozniak2000}.
The CTIO $V$ and $I$-band data were also reduced with DoPHOT 
\citep{Schechter1993} in order to get photometry of the lensed source on the
same scale as photometry of the non-variable bright stars in the frame.
Since there were only two observations from the
Faulkes Telescope South, we have not included these data in our modeling.
Finally, we did not use MOA data of before 2009, because there appeared to
be some systematic errors in the early baseline observations. 
The datasets used for the modeling are summarized in Table \ref{tab:dataset}.
The error bars provided by these photometry codes are generally good estimates of the
relative error bars for the different data points, but they often provide only
a rough estimate of the absolute uncertainty for each photometric measurement.
Therefore, we follow the standard practices of renormalizing the error bars 
to give $\chi^{2}/({\rm d.o.f.}) \sim 1$ for each data set once a reasonable model
has been found. In this case, we have used the best parallax plus orbital model
(see Section \ref{sec:parallax}) for this error bar renormalization.
This procedure ensures that the error bars for the microlensing fit parameters 
are calculated correctly. We carefully examined the property of residual distribution
weighted by the normalized errors. We confirmed that it is well represented by 
the Gaussian distribution with the sigma of close to unity, $\sigma=0.94$, where
Kolmogorov-Smirnov probability is 4.8\% and 6.9\% for the unconstrained 
Xallarap model and Parallax+orbital model, respectively (see next Section).
The best fit sigma of $\sigma=0.94$ is slightly smaller than unity due to compensating 
the excess points in the tails of the residual distribution.
The number of these excess points with more than 3$\sigma$ is not so large, $\sim$0.9 \%,
compared to formally expected fraction of 0.27\%.
Furthermore, they are sparsely distributed all over the light curve, i.e., not clustered at any particular place.
We also found that they are not correlated with seeing.  
There is only a weak correlation with airmass of $\sim$0.1 $\sigma$/airmass 
which is too small to explain the excess tails.
So we concluded that it is unlikely that they bias our result significantly.
Thus the effect of this small deviation from Gaussian was not tested in this work by 
more thorough analysis, like a bootstrap method. 

\section{Modeling}
\label{sec:modeling}

The parameters used for the standard binary lens modeling in this paper are
the time of closest approach to the barycenter of lens, $t_{0}$, the 
Einstein radius crossing time, $t_{\rm{E}}$, the impact parameter
in units of the Einstein radius, $u_{0}$, the planet-host mass ratio, 
$q$, the lens separation in the Einstein radius units, $s$, 
the angle of the source trajectory with respect to the binary axis, $\alpha$, 
and the angular source radius ($\theta_*$) normalized by the angular Einstein radius,
$\rho \equiv \theta_{*} / \theta_{\rm{E}}$.
The angular Einstein radius $\theta_{\rm{E}}$ is expressed as
$\theta_{\rm{E}} = \sqrt{\kappa M \pi_{\rm{rel}}}$, where $\kappa = 4G / \left( c^2 \rm{AU} \right) = 8.14 ~\rm{mas} ~\textit{M}^{-1}_{\sun}$,
and $M$ is the total mass of the lens system.
$\pi_{\rm{rel}}$ is the lens-source relative parallax given by 
$\pi_{\rm{rel}} = \pi_{\rm{L}} - \pi_{\rm{S}}$, 
where the $\pi_{\rm{L}} = \rm{AU}/D_{\rm{L}}$ and 
$\pi_{\rm{S}} = \rm{AU}/D_{\rm{S}}$ are the parallax of the lens 
and that of the source, respectively.
$D_{\rm{L}}$ is the distance to the lens, and $D_{\rm{S}}$ is that to the source.

First, we searched the standard model that minimizes $\chi^2$ with using the above parameters.
We used the Markov Chain Monte Carlo (MCMC) method to obtain the $\chi^{2}$ minimum.
Light curve calculations were done using a variation of the method of \citet{bennett-himag}.
The initial parameter-sets to search the standard model
were used over the wide range, $-5 \leq \log q \leq 0$ and $-1 \leq \log s \leq 1$.
The total number of initial parameter-sets was 858, and all parameters were free parameters.
We thereby found that the standard model showed $\chi^2 = 6038$,
and the parameters are listed in Table \ref{tab:parameter}.

\subsection{Limb darkening}

The caustic exit is well observed in this event, and this implies that 
finite source effects must be important, because caustic crossings
imply singularities in light curves for point sources.
We must therefore account for limb darkening when modeling this event.
We use a linear limb-darkening model in which the source 
surface brightness is expressed as 
\begin{equation}
  S_{\lambda}(\vartheta) = S_{\lambda}(0)[1 - u(1 - \cos \vartheta)] \ .
\end{equation}
The parameter $u$ is the limb-darkening coefficient, 
$S_{\lambda}(0)$ is the central surface brightness of the source, 
and the $\vartheta$ is the angle between the normal to the stellar surface and the line of the sight.
As discussed in Section \ref{sec:source},  the estimated intrinsic source color is 
$(V - I)_{\rm{S, 0}} = 0.70$, and its angular radius is $\theta_* = 0.91\pm 0.06\,\mu$as.
This implies that the source is mid-late G-type turn-off star.
From the source color, we estimate an effective temperature, 
$\rm{T}_{\rm{eff}} \sim 5690$ K according to \citet{Gonzalez2009}, 
adopting $\log[\rm{M/H}] = 0$. Assuming $\log g = 4.0 ~\rm{cm ~s}^{-2}$,
\citet{Claret2000} gives the limb-darkening coefficients for a  
$\rm{T}_{\rm{eff}} \sim 5750$ K star
of $u_{\rm{I}} = 0.5251$, $u_{\rm{V}} = 0.6832$, and 
$u_{\rm{R}} = 0.6075$, for $I$, $V$, and $R$-bands, respectively. 
We used the average of $I$ and $R$-band coefficient for MOA-Red wide band 
and $R$-band coefficient for unfiltered bands.

\subsection{Parallax}
\label{sec:parallax}

The orbital motion of the Earth during the event implies that the
lens does not appear to move at a constant velocity with respect
to the source, as seen by Earth-bound observers. This is known
as the (orbital) microlensing parallax effect, and it can often be detected
for events with time scales $t_{\rm{E}} > 50\,$days, like MOA-2010-BLG-328.
So, we have included this effect in our modeling.
This requires two additional parameters, $\pi_{\rm{E,N}}$ and $\pi_{\rm{E,E}}$, 
which are the two components of the microlensing
parallax vector $\mbox{\boldmath ${\pi}$}_{\rm{E}}$ \citep{Gould2000}.
The microlensing parallax amplitude is given by
$\pi_{\rm{E}} = \sqrt{\pi_{\rm{E,N}}^{2} + \pi_{\rm{E,E}}^{2}}$.
The amplitude $\pi_{\rm{E}}$ is also described as $\pi_{\rm{E}} = \pi_{\rm{rel}}/\theta_{\rm{E}}$.
The direction of $\mbox{\boldmath ${\pi}$}_{\rm{E}}$ is that of the lens-source relative proper motion 
at a fixed reference time of HJD=2455379.0 which is near the peak of event.
If both $\rho$ and $\pi_{\rm{E}}$ are measured,
one can determine the mass of the lens system,
\begin{equation}
  M = \frac{\theta_{\rm{E}}}{\kappa \pi_{\rm{E}}} 
    = \frac{\theta_{*}}{\kappa \rho \pi_{\rm{E}}}\ ,
  \label{eq:mass}
\end{equation}
assuming one also has an estimate of $\theta_*$, the angular source radius.
Since the source distance, $D_{\rm S} = \rm{AU}/\pi_{\rm S}$, is approximately known, 
we can also estimate the lens distance from
\begin{equation}
  D_{\rm L} = \frac{{\rm AU}}{\pi_{\rm E}\theta_{\rm E}+\pi_{\rm S}}
                = \frac{{\rm AU}}{\pi_{\rm E}\theta_{*}/\rho + \pi_{\rm S}} \ .
  \label{eq-Dl}
\end{equation}

The parallax model parameters are shown in Table \ref{tab:parameter}, and as this
table indicates, inclusion of the parallax parameters improves $\chi^{2}$
by $\Delta \chi^{2} = 353$. When the parallax effect is relatively weak, there is
an approximate symmetry in which the lens plane is replaced by its mirror
image (i.e.\ $u_0 \rightarrow - u_0$ and $\alpha \rightarrow - \alpha$). However,
for this event, this symmetry is broken as
the $u_{0} > 0$ solution yields a $\chi^{2}$ smaller than the $u_{0} < 0$ solution 
by $\Delta \chi^{2} = 78$.

\subsection{Orbital motion of the lens companion}

Another higher order effect that is always present is the orbital motion 
of the lens system. This causes the shape and position of the caustic curves
to change with time. The microlensing signal of the planet can be seen
for only $\sim 5\,$days, which is much smaller than the likely orbital period
of $\sim 8\,$years, so it is sensible to consider the lowest order components
of orbital motion, the two-dimensional relative velocity in the plane of the sky.
To lowest order, orbital motion can be expressed by velocity components
in polar coordinates, $\omega$ and $ds/dt$. These are the binary rotation 
rate and the binary separation velocity \citep{Dong2009}. (Note that this
would be a poor choice of variables in cases (e.g., \citealt{Bennett2010}) where
the binary acceleration is important, because polar coordinates are not
inertial.) When written in the (rotating) lens coordinate system,
the source trajectory takes the form
\begin{eqnarray}
  \alpha(t) &=& \alpha_{0} + \omega ( t - t_{0}) ~, \\
       s(t) &=& s_{0} + ds/dt ( t - t_{0}) ~.
\end{eqnarray}

\begin{figure}[htbp]
 \vspace{35mm}
  \includegraphics[angle=-90,scale=0.65,keepaspectratio]{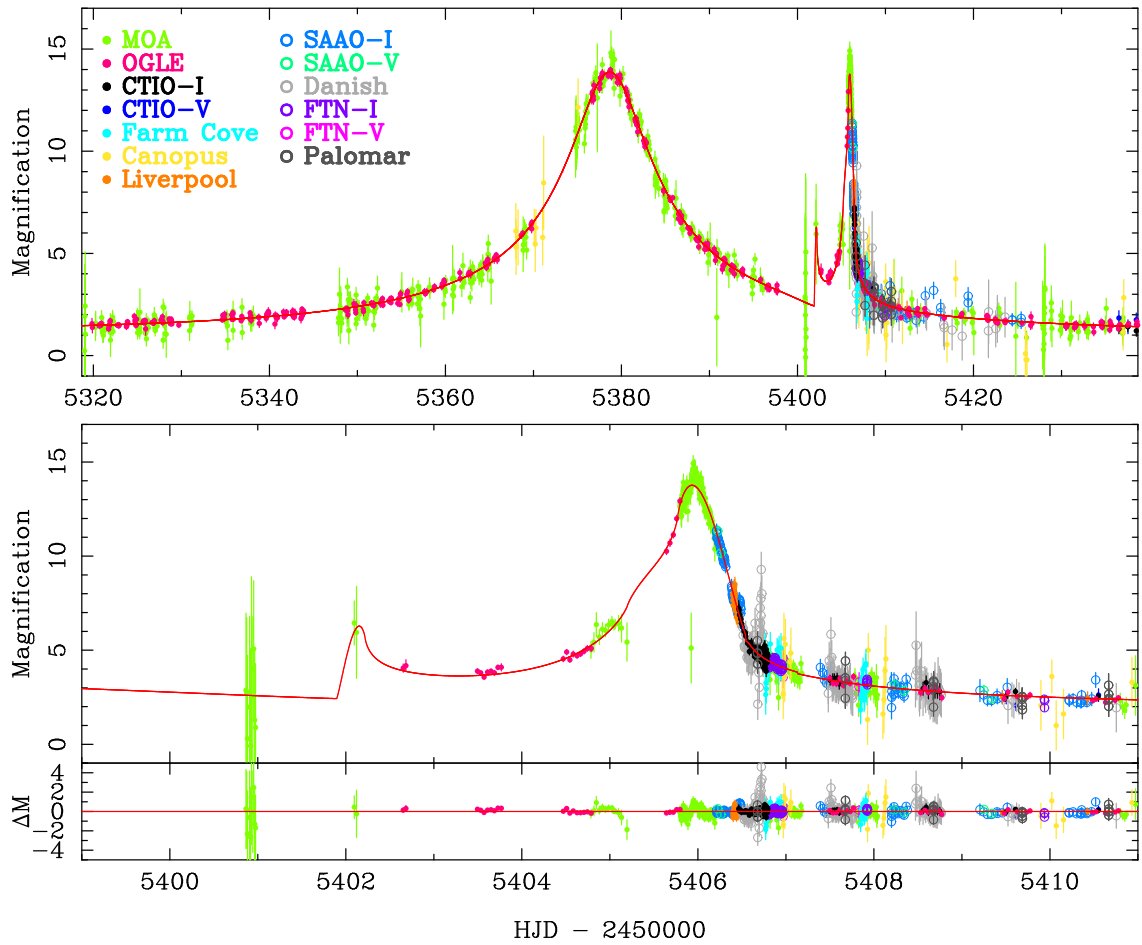}
 \caption{The best parallax plus orbital motion model light curve :
          The top panel shows the whole lightcurve and the middle panel shows anomaly.
          The bottom panel indicates residuals from the model.}
 \label{fig:lightcurve}
\end{figure}

\begin{figure}[htbp]
 \epsscale{0.9}
 \includegraphics[angle=-90,scale=0.65,keepaspectratio]{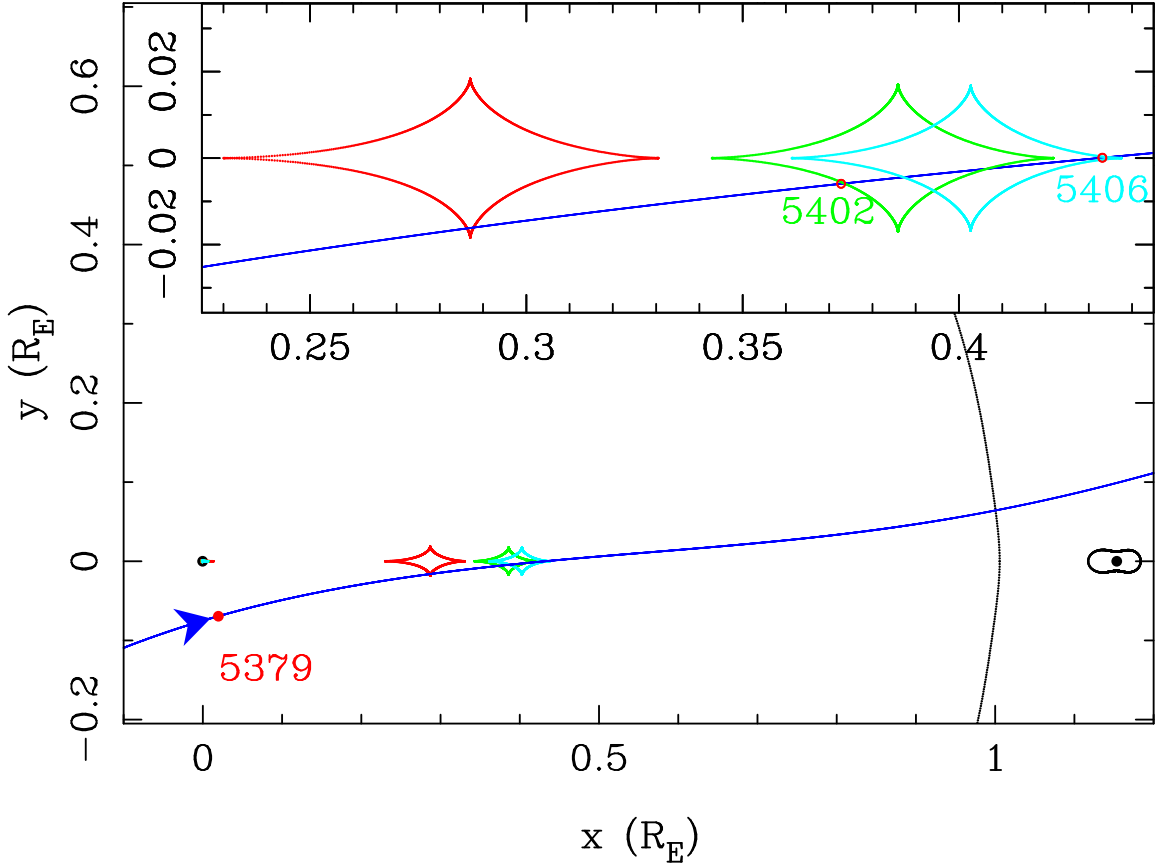}
 \caption{Caustics (red lines) and critical curves (black lines) of 
 the best parallax plus orbital motion model near the peak at HJD=2355379. 
 The source trajectory is shown by the blue lines. 
 The black dot $(x, y) \sim (1.2, 0)$ represents the planet position.
 Green and cyan lines indicate the caustics when the source enter the 
 caustic at HJD=2355402 and exit at HJD=2355406.
 The inset shows a closeup of the planetary caustic.
 The red filled and open circles on the source trajectory are source positions at 
 HJD=2355379, 2355402 and 2355406, respectively.
 The size of the red open circles in the inset indicates the source size.}
 \label{fig:caustic}
\end{figure}

We have conducted fits with both orbital motion alone and with microlensing
parallax plus orbital motion, and the best fit parameters for each model are given
in Table \ref{tab:parameter}. 
Figure \ref{fig:lightcurve} presents the light curve of the best parallax plus orbital motion model, and Figure \ref{fig:caustic} shows its caustic.
This table indicates that the orbital motion only 
model improves $\chi^2$ by $\Delta\chi^2 = 322$ vs.\ the standard model,
which is slightly worse than the $\chi^2$ improvement of $\Delta\chi^2 = 353$
for the parallax only model.
The combined parallax and orbital motion
model yield a $\chi^2$ improvement of $\Delta\chi^2 = 373$.
As shown in Table~\ref{tab:parameter}, we found that
the $u_0 < 0$ model showed smaller $\chi^2$ than the $u_0 > 0$ models
for the parallax plus orbital motion model, but
the difference between $\chi^2$ of the $u_0 < 0$ model and that of the $u_0 > 0$ model is small ($\Delta \chi^2 \sim 3$).
This is due to the degeneracy of $\pi_{\rm{E}, \perp}$ with $\omega$ \citep{Batista2011, Skowron2011}, 
where $\pi_{\rm{E}, \perp}$ is the component of $\mbox{\boldmath ${\pi}$}_{\rm{E}}$ that is perpendicular
to the instantaneous direction of the Earth's acceleration.
Compared to the $u_0 > 0$ model, 
the $u_0 < 0$ model is preferred by $\Delta \chi^2 \sim 3$ but the $u_0 > 0$ model cannot be excluded.

Since both orbital motion and microlensing parallax 
should exist at some level in every binary microlensing light curve, the
parallax plus orbital motion model should be considered more realistic
than the orbital motion only model.
However, it is important to check that the parameters of the orbital motion
models are consistent with the allowed velocities for bound orbits, since the
probability of finding planets in unbound orbits is extremely small.
So we would like to be able to compare the transverse kinetic energy, $\left(\rm{KE} \right)_{\perp} = M v_{\rm{rel}, \perp}^2/2$,
with the potential energy, $\left(\rm{PE} \right)_{\perp} = G M/r_{\perp}$ \citep{Dong2009}.
Then, the ratio of kinetic to potential energy can be expressed in terms of observables as
\begin{equation}
  \left( \frac{\rm{KE}}{\rm{PE}} \right)_{\perp} =
  \frac{\kappa M_{\sun} \pi_{\rm{E}} \left( |\mbox{\boldmath ${\gamma}$}| \rm{yr} \right)^2 s_{0}^3}
       {8 \pi^2 \theta_{\rm{E}} \left( \pi_{\rm{E}} + \pi_{\rm{S}}/\theta_{\rm{E}} \right)^3 } ~,
\end{equation}
where $\mbox{\boldmath ${\gamma}$} = (\gamma_\parallel, \gamma_\perp)$ consists of $\gamma_\parallel = (ds/dt)/s_0$ and $\gamma_\perp = \omega$.
The parameters of the parallax plus orbital motion model indicate that
these ratio are 0.72 and 0.08 for the $u_0<0$ and $u_0>0$ model, respectively, 
and this implies that the both models are reasonable.

\subsection{Xallarap} 
\label{sec:xallarap}

The xallarap effect is the converse of the parallax effect.
It is due to the orbital motion of the source instead of the orbital motion of the observers on the Earth.
Xallarap can cause similar light curve distortions to the parallax effect
\citep{Poindexter2005}. Unlike parallax and orbital motion, however, there
is a good chance that the source will not have a companion with an
orbital period in the right range to give a detectable xallarap signal. 
Only about 10\% of source stars have a companion with orbital
parameters that would allow a xallarap solution that could mimic
microlensing parallax.

For the xallarap model, the xallarap vector, ($\xi_{\rm{E,N}}$, $\xi_{\rm{E,E}}$), 
which correspond to ($\pi_{\rm{E,N}}$, $\pi_{\rm{E,E}}$), 
the direction of observer relative to the source orbital axis, $\rm{R.A.}_{\xi}$ and $\rm{decl.}_{\xi}$, 
the orbital period, $P_{\xi}$, the orbital eccentricity, $\epsilon$, 
and the time of periastron, $t_{\rm{peri}}$, are required in addition to the standard binary model.

The xallarap amplitude, $\xi_{\rm{E}}$, is expressed with Kepler's third law as follows 
\begin{equation}
  \xi_{\rm{E}} = \frac{a_{\rm{s}}}{\hat{r}_{\rm{E}}} = \frac{1 ~\rm{AU}}{\hat{r}_{\rm{E}}} \biggl( \frac{M_{\rm{c}}}{M_{\sun}} \biggr) \biggl(\frac{M_{\sun}}{M_{\rm{c}} + M_{\rm{S}}} \frac{P_{\xi}}{1 ~\rm{yr}} \biggr)^{\frac{2}{3}},
\end{equation} 
where $a_{\rm{s}}$ is the semimajor axis of the source orbit and, $M_{\rm{S}}$ and $M_{\rm{c}}$ 
are the masses of the source and its companion, respectively. 
The $\hat{r}_{\rm{E}}$ is the Einstein radius projected on the source plane and is described as follows
\begin{equation}
  \frac{\hat{r}_{\rm{E}}}{\rm{AU}} = \theta_{\rm{E}} D_{\rm{S}} = \frac{\theta_{*}}{\rho} D_{\rm{S}} ~.
\end{equation} 
Assuming values for the two masses $M_{\rm{S}}$ and $M_{\rm{c}}$,
we can determine the $\xi_{\rm{E}}$ for a given period $P_{\xi}$,
and then we can constrain the $\xi_{\rm{E}}$ value in the xallarap model.
We conducted the xallarap modeling with constraint and without constraint.
For the constrained xallarap model, we assumed  $M_{\rm{S}} =1 M_{\sun}$, $D_{\rm{S}} = 8 ~\rm{kpc}$,
and various masses of companion, $M_{\rm{c}}$, from $0.1 M_{\sun}$ to $1 M_{\sun}$.
Here the upper limit of the companion mass is due to the measured blending flux
as shown in Section \ref{sec:ConstrainedXallarap}.

\begin{figure}[htbp]
 \epsscale{0.8}
 \plotone{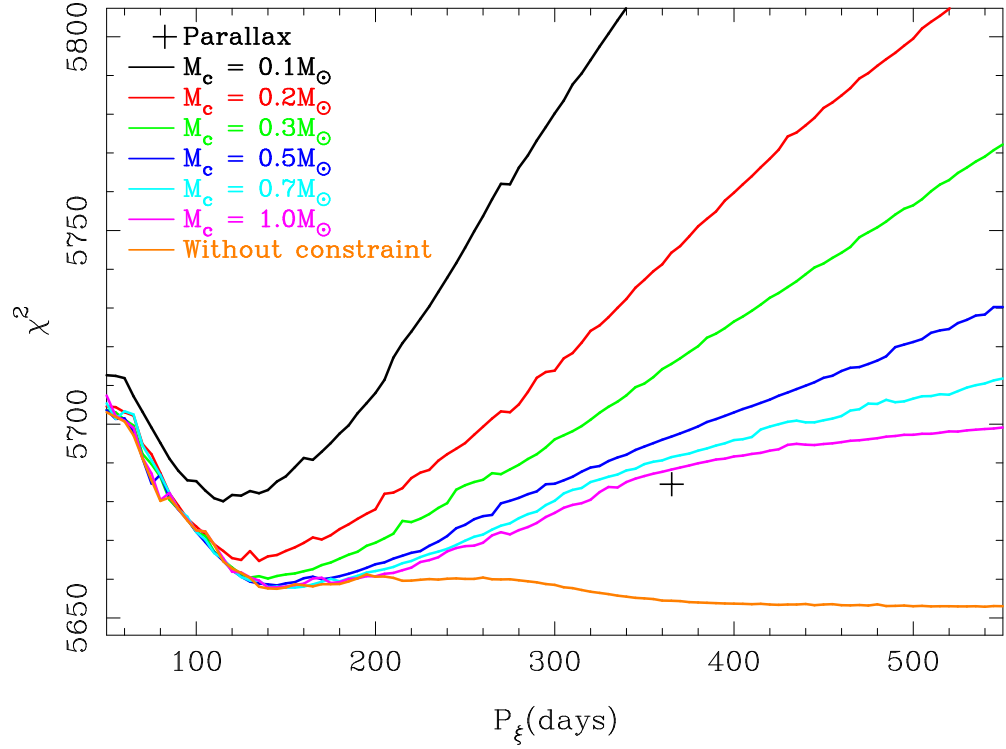}
 \caption{The $\chi^{2}$ of the xallarap model as a function of the orbital period. 
 The color-coded lines represent the unconstrained model and 
 the constrained model assuming various companion masses, respectively.
 The ``+" indicates the best parallax model (with no orbital motion) for comparison.}
 \label{fig:chi2}
\end{figure}

The parameters obtained for each models are listed in Table \ref{tab:parameter}.
At this time, we ignored the orbital motion of the companion of the lens. 
The best unconstrained and constrained xallarap models (fixed with $M_{\rm{c}} = 1 M_{\sun}$) have nearly identical $\chi^2$,
i.e., 5652 and 5653, respectively.
They gave improved with $\Delta \chi^{2} \sim 385$ compared with the standard model
and $\Delta \chi^{2} \sim 5$, $6$ compared with the parallax plus orbital motion model.
Figure \ref{fig:chi2} shows the $\chi^{2}$ distribution as a function of $P_{\xi}$.
In Figure \ref{fig:chi2}, the orbital eccentricity, $\epsilon$, was fixed as the value for Earth.
Only the result with companion mass $M_{\rm{c}} = 0.1 M_{\sun}$ shows worse $\chi^{2}$ 
than the parallax only model for every $P_{\xi}$.
Therefore we estimated the probability of the existence of the companion with $0.2 < q < 1$ and 
$80 < P_{\xi} < 365$, whose $\chi^{2}$ were smaller than parallax model, 
and found that the prior probability was about 3 \% \citep{Duquennoy1991}.

\begin{figure}[htbp]
 \epsscale{0.8}
 \plotone{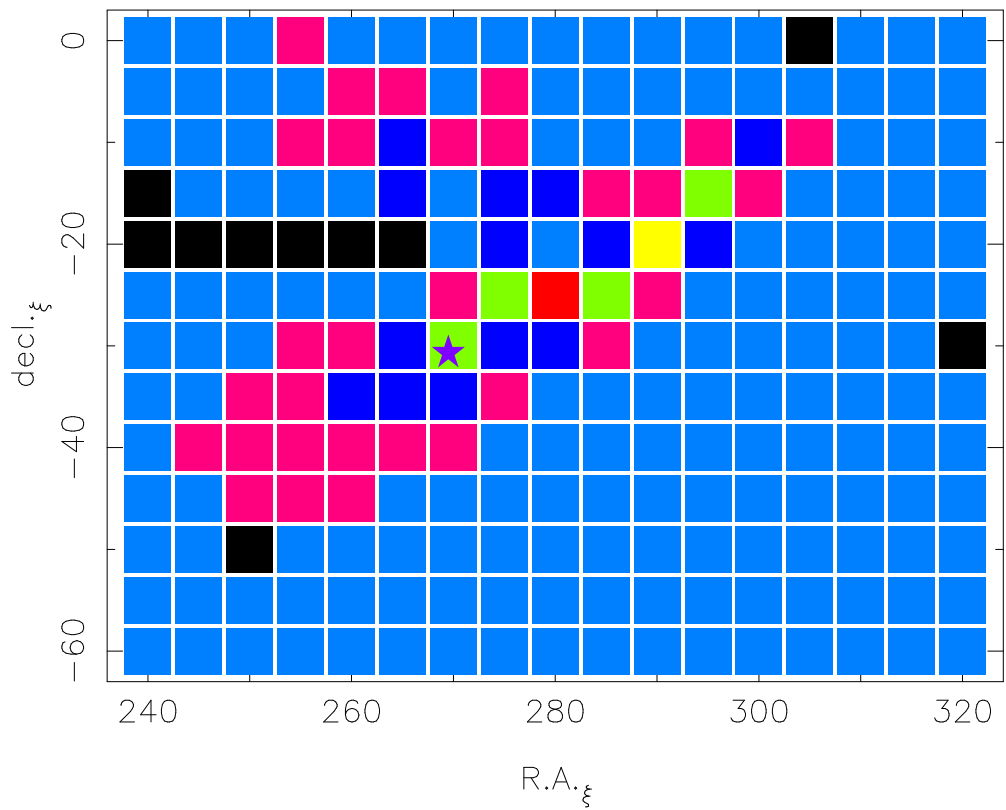}
 \caption{A map of $\chi^{2}$ surface of the unconstrained xallarap model with fixed $\rm{R.A.}_{\xi}$ and $\rm{decl.}_{\xi}$ and $u_0>0$.
 The period and eccentricity are fixed at those of the Earth.
 The orbital motion of the lens is included.
 The square is color-coded for solutions with $\Delta \chi^2$ within 1(red), 4(yellow), 9(green), 16(blue),
 25(magenta), 49(aqua), and the black indicates larger than 49.
 The purple star mark represents the position of the target in the sky plane ($\rm{R.A.} = 269^{\circ}$, $\rm{decl.} = -31^{\circ}$).}
 \label{fig:map}
\end{figure}

As mentioned earlier in this section, the xallarap signal can mimic the microlening parallax.
If the parallax signal is real, the xallarap parameters should converge on the Earth values.
To check whether the parallax is real, we verify the parameters of the unconstrained xallarap model.
This is because if the parallax model is correct,
the xallarap parameter, $\xi_{\rm{E}}$, does not represent a real companion, so it can be anything.
Focusing on the period, $P_{\xi}$, as shown in Figure \ref{fig:chi2}, the period is consistent with $P_{\xi} = 1 \rm{yr}$.
Note that the lens orbital motion is ignored in Figure \ref{fig:chi2}.
Then we checked the consistency of the $\rm{R.A.}_{\xi}$ and $\rm{decl.}_{\xi}$.
To check this, we conducted the xallarap modeling with $\rm{R.A.}_{\xi}$ and $\rm{decl.}_{\xi}$ fixed at grid of values.
During this test, we fixed the eccentricity and period with the Earth's values and included the lens orbital motion, i.e., $\omega$, $ds/dt$. 
The $\chi^2$ map in the $\rm{RA}_{\xi}$-$\rm{decl}_\xi$ plane is shown in Figure \ref{fig:map}.
The best xallarap model has ($\rm{R.A.}_{\xi}$, $\rm{decl.}_{\xi}$) = (280, -25) 
that is close to the coordinate of the event ($\rm{R.A.}$ = 269$^{\circ}$, $\rm{decl.}$ = -31$^{\circ}$).
The $\Delta \chi^2$ between best xallarap model and nearby coordinate of the event ($\rm{R.A.}$ = 270$^{\circ}$, $\rm{decl.}$ = -30$^{\circ}$) is small ($\Delta \chi^2\sim 7$).
The results of the verification of the consistency of the period and coordinate support that
the xallarap parameters are consistent with the Earth parameters.
As mentioned before, the prior probability that the source has a
companion with the required mass/period parameters is small ($\sim 3\%$).
Now, even if it did have these parameters, the chance that the
orientation of the orbit would mimic that of the Earth's to the degree
shown in Figure 4 is only about 0.6\%.  Combining these two factors
yields a prior probability, $B_{\rm{x}}$, of only $2\times 10^{-4}$.
On the other hand, we also estimated a prior probability that the lens would be at 0.81 kpc
(as derived from the parallax solution), $B_{\rm{p}}$, at 2.5\%.
This is 2$\sigma$ value of the MCMC chain of the parallax plus orbital motion.
Note that the distribution of the error is not a gaussian.
Comparing both prior probabilities, the parallax is preferred by a factor of $B_{\rm{p}}/B_{\rm{x}} = 125$.
Moreover we found that the probability of getting the observed improvement in $\Delta \chi^2$ with the
xallarap model for additional degrees of freedom, $P(\Delta \chi^2; \rm{N})$, was about 0.1.
Even if we consider this probability, the parallax solution is still preferred a factor of $B_{\rm{p}}/B_{\rm{x}} \times P(\Delta \chi^2; \rm{N}) = 12.5$.
Nevertheless, the prior probability that the lens would be at 0.81 or 1.24 kpc (as derived from the parallax solution) is also relatively
low, so the xallarap model needs to be considered carefully.
We therefore estimated the physical parameters of the lens each for the parallax plus orbital motion model
and for the constrained xallarap model by a Bayesian analysis with using $t_{\rm{E}}$ and $\theta_{\rm{E}}$.

\section{Source star properties}
\label{sec:source}

\subsection{For parallax plus orbital motion model}
We derive the source star angular radius, $\theta_{*}$,
in order to obtain the angular Einstein radius, $\theta_{\rm{E}}$.
The model-independent source color is derived from $V$ and $I$-band photometry of CTIO using linear regression,
and the observed magnitude of the source is determined from OGLE $I$-band with modeling.
However, these are affected by interstellar dust.
This means that we need to estimate the intrinsic source color and magnitude.
Therefore we use the red clump giants (RCG), which are known to be approximate standard candles.
We adopt the intrinsic RCG color $(V - I)_{\rm{RCG, 0}} = 1.06 \pm 0.06$ \citep{Bensby2011}
and the magnitude $M_{I, \rm{RCG}, 0} = 14.45 \pm 0.04$ \citep{Nataf2012},
\begin{equation}
  (V - I, I)_{\rm{RCG, 0}} = (1.06, 14.45) \pm (0.06, 0.04).
\end{equation}

\begin{figure}[htbp]
 \epsscale{0.75}
 \plotone{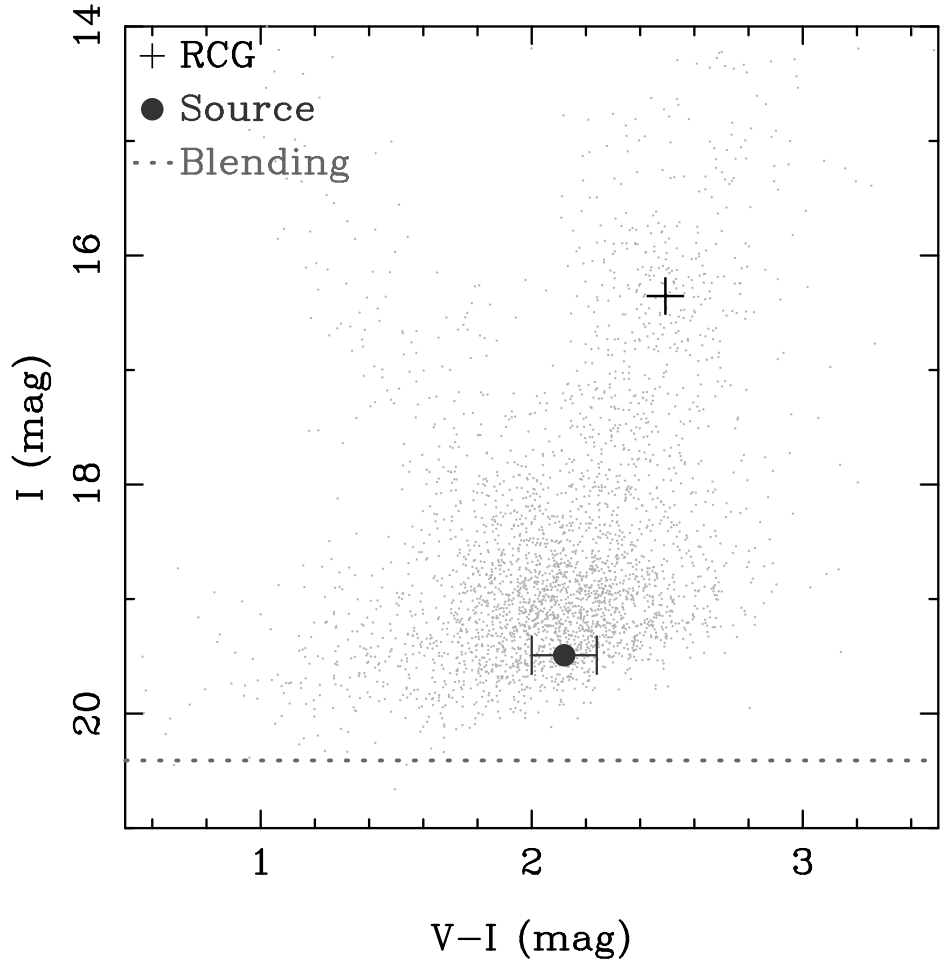}
 \caption{The OGLE-III color-magnitude diagram of stars within 1.5' from the source star of MOA-2010-BLG-328.
 The filled circle represents the $I$-band magnitude of the source and the horizontal dashed lines indicates the blended light in the best parallax plus orbital motion.
 The cross indicates the center of Red Clump Giants.}
 \label{fig:CMD}
\end{figure}

We construct two color-magnitude diagrams (CMD): one is constructed of the OGLE-III catalog
and the other is constructed of the instrumental CTIO photometry.
The CMD of OGLE-III is used for the calibration of the $I$-band magnitude,
and the CMD of CTIO is used for the calibration of the color.
Figure \ref{fig:CMD} shows the CMD constructed of the OGLE-III catalog.
From the CMD, the $I$-band magnitude of the RCG centroid is estimated to be $I_{\rm{RCG, obs}} = 16.28$.
By comparing the intrinsic RCG and observed one, we find that the offset is $\Delta I_{\rm{RCG}} = 1.83$.
Applying this offset to the observed $I$-band magnitude of the source,
$I_{\rm{S, obs}} = 19.49$ ($u_0 < 0$ model),
we obtain the intrinsic $I$-band magnitude of the source, $I_{\rm{S, 0}} = 17.66$.
Likewise, we estimate that the instrumental color of the RCG centroid of CTIO CMD is $(V - I)_{\rm{RCG, obs}} = 0.80$.
The color offset of RCG between the intrinsic and the observed value is $\Delta (V - I)_{\rm{RCG}} = 0.26$.
Using this offset to calibrate the observed color of the source, $(V - I)_{\rm{S, obs}} = 0.44$, 
we can get the intrinsic color of the source as $(V - I)_{\rm{S, 0}} = 0.70$.
Finally we find the intrinsic $I$-band magnitude and the color of the source is
\begin{equation}
  (V - I, I)_{\rm{S, 0}} = (0.70, 17.66) \pm (0.10, 0.04) ~(u_0 < 0) ~.
\end{equation}
Using color-color relation \citep{Bessell1988}, we derive $(V - K, K)_{\rm{S, 0}}$ from $(V - I, I)_{\rm{S}, 0}$,
\begin{equation}
  (V - K, K)_{\rm{S, 0}} = (1.51, 16.89) \pm (0.23, 0.26) ~(u_0 < 0) ~.
\end{equation}
Then, we apply the relation between $(V-K, K)_{\rm{S, 0}}$ and the stellar angular radius \citep{Kervella2004}
and estimate the source star angular radius, 
\begin{equation}
  \theta_{*} = 0.91 \pm 0.06 ~\mu \rm{as} ~(u_0 < 0) ~.
\end{equation}
Adopting same procedure for $u_0 > 0$ model, we derive $\theta_{*} = 0.90 \pm 0.06 ~\mu \rm{as}$.
This source star angular radius is consistent with that obtained from $u_0 < 0$ model.
These source star angular radii mean that the source star radius is 1.5 $ R_{\sun}$ 
with assuming the source star locates in the Galactic bulge ($\sim$ 8 kpc).
The color of the source star indicates that the source star is G-star,
and the estimated source star radius is slightly larger than typical G-dwarfs.
For this reason, we conclude that the source star is G-subgiant or turn-off star.
From the finite source effect parameter, $\rho$, in the parallax plus orbital motion model, 
we drive the angular Einstein radius and source-lens relative proper motion, $\mu$, for $u_0 < 0$ model
\begin{eqnarray}
  \theta_{\rm{E}} &=& \frac{\theta_{*}}{\rho} = 0.98 \pm 0.12 ~\rm{mas} ~, \\
   \mu &=& \frac{\theta_{\rm{E}}}{t_{\rm{E}}} = 5.71 \pm 0.70 ~\rm{mas/yr} ~,
\end{eqnarray}
and for $u_0 > 0$ model,
\begin{eqnarray}
  \theta_{\rm{E}} &=& 0.83 \pm 0.14 ~\rm{mas} ~, \\
                \mu &=& 4.72 \pm 0.79 ~\rm{mas/yr} ~.
\end{eqnarray}

\subsection{For constrained xallarap model}

According to the same procedure used for the parallax plus orbital motion model,
we also estimated source star properties for the case of the constrained xallarap model.
The source color and magnitude are largely similar to
those that are obtained from the parallax plus orbital motion model,
\begin{eqnarray}
  (V - I, I)_{\rm{S, 0}} &=& (0.70, 17.63) \pm (0.10, 0.04) ~, \\
  (V - K, K)_{\rm{S, 0}} &=& (1.51, 16.83) \pm (0.23, 0.26) ~.
\end{eqnarray}
From these source color and magnitude,
we derived the angular Einstein radius and source-lens relative proper motion, 
\begin{eqnarray}
  \theta_{\rm{E}} &=& 0.68 \pm 0.04 ~\rm{mas} ~, \\
              \mu &=& 4.03 \pm 0.26 ~\rm{mas/yr} ~.
\end{eqnarray}

\section{Lens system}
\label{sec:lens}

\subsection{Parallax plus orbital motion model}

For determining the mass and distance of lens system, 
we combine Equations (\ref{eq:mass}), (\ref{eq-Dl}) and the microlensing parallax parameter,
$\pi_{\rm{E}}$, which was derived from the parallax plus orbital motion model.
For the $u_0 < 0$ model, Equation (\ref{eq:mass}) yields a host star mass of 
$M_{\rm{h}} = 0.11 \pm  0.01 M_{\sun}$, and a planet mass of 
$M_{\rm{p}} = 9.2 \pm 2.2 M_{\oplus}$. 
To determine the distance to the lens system with equation  (\ref{eq-Dl}),
we need the source distance, $D_{\rm S}$,  which we assume to be  $D_{\rm S} = 8.0 \pm 0.3$ kpc \citep{Yelda2010},
i.e. $\pi_{\rm{S}} = 0.125 \pm 0.005$ mas, and this gives a lens distance of
$D_{\rm{L}} = 0.81 \pm 0.10 ~\rm{kpc}$. 
The projected star-planet separation is therefore $r_{\perp} = s D_{\rm{L}} \theta_{\rm{E}} = 0.92 \pm 0.16 ~\rm{AU}$.
These implies that the lens is very nearby red star.
The probability distributions of the mass, distance, Einstein radius, and brightnesses of the lens are shown in Figure \ref{fig:bayesian}.

The mass of host star, derived from the parallax plus orbital motion model,
indicates that the absolute $J$, $H$, and $K$-band magnitudes of that would be
$M_J = 10.06 \pm 0.29$, $M_H = 9.49 \pm 0.27$, and $M_K = 9.16 \pm 0.25$ mag, 
respectively \citep{Kroupa1997}.
\citet{Marshall2006} calculated the extinction distribution in three dimensions.
According to them, the extinction in $K$ band at $D_{\rm{L}} = 0.81 \pm 0.10$ kpc 
is $A_K = 0.05 \pm 0.01$. The \citet{Cardelli1989} extinction law gives
infrared extinction ratios of $A_J:A_H:A_K = 1:0.67:0.40$, which implies that
the $J$ and $H$-band extinctions are $0.13 \pm 0.02$ and $0.09 \pm 0.01$.
With these extinctions and the derived distance modulus,
the apparent $J$, $H$, and $K$-magnitudes of the host (and lens) star would be
$J_{\rm L} = 19.73 \pm 0.39$, $H_{\rm L} = 19.13 \pm 0.37$, and 
$K_{\rm L} = 18.76 \pm 0.36$ mag, respectively.

For the $u_0 > 0$ model, we find that a host star mass of 
$M_{\rm{h}} = 0.12 \pm  0.02 M_{\sun}$, a planet mass of 
$M_{\rm{p}} = 15.2 \pm 5.9 M_{\oplus}$, a lens distance of
$D_{\rm{L}} = 1.24 \pm 0.18 ~\rm{kpc}$. 
The projected star-planet separation is therefore $r_{\perp} = 1.21 \pm 0.27 ~\rm{AU}$.
The mass of host star indicates that the absolute $J$, $H$, and $K$-band magnitudes of that would be
$M_J = 9.74 \pm 0.38$, $M_H = 9.19 \pm 0.36$, and $M_K = 8.88 \pm 0.34$ mag, respectively.
The distance to the lens indicates that the extinction in $J$, $H$, and $K$-band
are $A_J = 0.20 \pm 0.03$, $A_H = 0.14 \pm 0.02$, and $A_K = 0.08 \pm 0.01$.
With these extinctions and the derived distance modulus,
the apparent $J$, $H$, and $K$-magnitudes of the host (and lens) star would be
$J_{\rm L} = 20.40 \pm 0.50$, $H_{\rm L} = 19.80 \pm 0.48$, and 
$K_{\rm L} = 19.42 \pm 0.47$ mag, respectively.

\subsection{Constrained xallarap model}
\label{sec:ConstrainedXallarap}

For the xallarap model, we estimate lens properties using a Bayesian analysis. 
We can obtain only the Einstein angular radius, $\theta_{\rm{E}}$, 
from the finite source effect parameter, $\rho$, in xallarap model. 
Consequently, for a Bayesian analysis, we combined Equation (\ref{eq:mass}), (\ref{eq-Dl}) 
and $\theta_{\rm{E}}$, $t_{\rm{E}}$ with the Galactic model \citep{Han2003}, and mass function.

The mass function is based on \citet{Sumi2011} Table 3S model \#1, but we apply a slight modification.
\citet{Sumi2011} assumed that stars that were initially above 1$M_{\sun}$ have evolved 
into stellar remnants.
However we assume the fraction of stars with mass of above 1$M_{\sun}$ by reference to \citet{Bensby2011}.
\citet{Bensby2011} obtained spectra of 26 microlensed stars and found that 12 stars were old metal poor, and
14 stars were young metal rich stars.
So we assume that the mass function is constructed both by old metal poor stars and young metal rich stars equally.
From the isochrones of \citet{Demarque2004}, we conclude that old metal poor stars, 
the initial mass of which were above 1$M_{\sun}$, have evolved into stellar remnants and
young metal rich stars, the initial mass of which were above 1.2$M_{\sun}$, 
have evolved into stellar remnants.
Hence, the mass function has a half fraction of initial mass function
above 1$M_{\sun}$ and has a cutoff at 1.2$M_{\sun}$.

Additionally, we used the $I$-band blended magnitude as an upper limit. 
The blended light derived from the modeling is $I_{\rm{b, obs}} = 20.49 \pm 0.11$.
Even if the lens lies behind all the dust, it cannot have higher dereddened light than
the light to which the same offset as the source is applied, $I_{\rm{b, 0}} = 18.66 \pm 0.12$.
Thereby we used this blended light as an upper limit in a Bayesian analysis.

\begin{figure}[htbp]
 \epsscale{0.75}
 \plotone{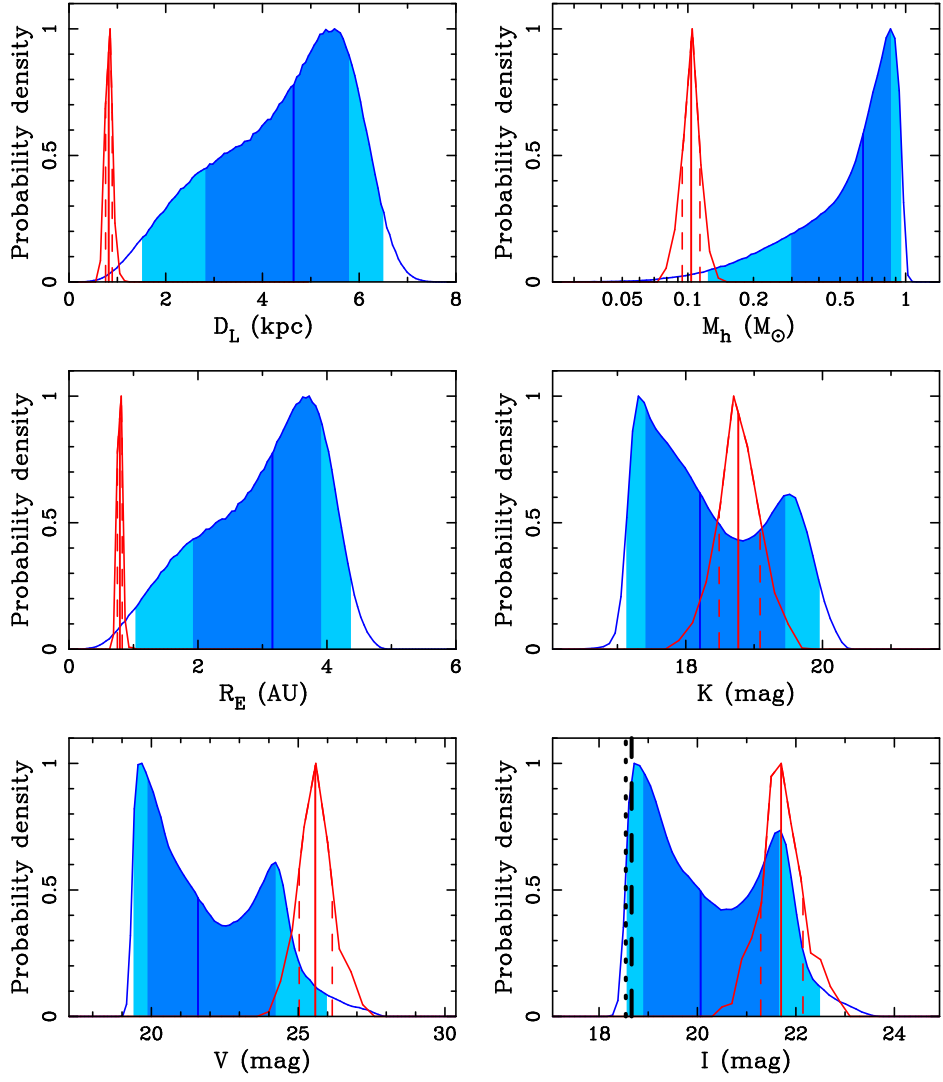}
 \caption{
The blue distributions are the probability distributions of the distance to the lens, $D_{\rm{L}}$,
mass of the host star, $M_{\rm{h}}$, Einstein radius, $R_{\rm{E}}$,
$K$, $V$ and $I$-band magnitudes for the best constrained xallarap model.
The vertical solid lines indicate the median values.
The dark and light shaded regions indicate the 1$\sigma$ and 2$\sigma$ limits.
The vertical dashed and dotted lines in the $V$ and $I$-band panels express the observed
upper limit and $1 \sigma$ error.
The right peaks of the distributions of the magnitude consist of nearby red (low mass) stars, and the left peaks consist of far blue (massive) stars.
The red distributions are the probability distributions estimated by MCMC chains of the best parallax plus orbital model for comparison.
The vertical solid and dashed lines indicate the median values and the $1 \sigma$ limits.
}
 \label{fig:bayesian}
\end{figure}

Figure \ref{fig:bayesian} shows the probability distributions obtained by a Bayesian analysis.
From the analysis, we find that the host star is a K-dwarf with mass  
$M_{\rm{h}} = 0.64_{-0.34}^{+0.22} M_\sun$,
and distance $D_{\rm{L}} = 4.6_{-1.8}^{+1.1}$ kpc.
The planet has a Saturn-like mass, 
$M_{\rm{p}} = 109_{-58}^{+38} M_\oplus = 1.15_{-0.61}^{+0.40} M_{\rm{Saturn}}$.
The Einstein radius is $R_{\rm{E}} = 3.2_{-1.2}^{+0.8}$, implying projected separation is
$r_\perp   = 3.8_{-1.5}^{+0.9}$ AU.
The physical three-dimensional separation is $a = 4.6_{-1.7}^{+1.9}$ AU, 
estimated by putting a planetary orbit at random inclination and phase \citep{Gould1992}.

The Bayesian analysis also yields the $J$, $H$, and $K$-band magnitudes of the host star,
which are $J_{\rm L0}  = 18.95^{+1.32}_{-0.99}$, $H_{\rm L0} =  18.37^{+1.33}_{-0.85}$, and
$K_{\rm L0} =  18.21^{+1.25}_{-0.80}$ mag, respectively, without extinction.
In Figure \ref{fig:bayesian}, the distributions of the magnitude have two peaks.
The right peak consists of nearby red (low mass) stars, and the left peak consists of far blue (massive) stars.
The distance to the lens indicates the extinctions of $A_J = 0.72^{+0.01}_{-0.25}$, 
$A_H = 0.49^{+0.01}_{-0.17}$, and $A_K = 0.29^{+0.01}_{-0.10}$.
According to these estimates, the apparent magnitudes of the host (and lens)
star should be
$J_{\rm L} = 19.67^{+1.32}_{-1.02}$, $H_{\rm L} = 18.86^{+1.33}_{-0.87}$, and 
$K_{\rm L} = 18.50^{+1.25}_{-0.81}$ mag.

\section{Discussion and conclusion}
\label{sec:discussion}

We report the analysis of the planetary microlensing event MOA-2010-BLG-328. 
The higher order effect improved the $\chi^{2}$ and the constrained xallarap model yielded the smallest $\chi^{2}$ value. 
However the difference of the $\chi^{2}$ between the constrained xallarap and 
the parallax plus lens orbital motion model is small ($\Delta \chi^{2} = 5$), 
and the xallarap has a high probability to mimic the parallax for this event.
We found that the mass ratio and separation are $(2.60 \pm 0.53) \times 10^{-4}$ and
$1.154 \pm 0.016$ Einstein radii for the best $u_0 < 0$ parallax plus orbital model,
$(3.68 \pm 1.26) \times 10^{-4}$ and
$1.180 \pm 0.028$ Einstein radii for the best $u_0 > 0$ parallax plus orbital model,
and
$(5.17 \pm 0.08) \times 10^{-4}$ and $1.216 \pm 0.001$ Einstein radii for the best constrained xallarap model.

Using parallax parameter $\pi_{\rm{E}}$, we can determine the physical parameters of the lens uniquely.
In the case of $u_0 < 0$ model, the mass of the host star and distance to the lens are derived to 
$M = 0.11 \pm 0.01 M_{\sun}$ and $D_{\rm{L}} = 0.81 \pm 0.10$ kpc.
The mass of the planet is $M_{\rm{p}} = 9.2 \pm 2.2 M_{\oplus}$ and projected separation is 
$r_{\perp} = 0.92 \pm 0.16$ AU.
On the other hand, in the case of $u_0 > 0$ model, the mass of the host star and distance to the lens are derived to
$M = 0.12 \pm 0.02 M_{\sun}$ and $D_{\rm{L}} = 1.24 \pm 0.18$ kpc.
The mass of planet is $M_{\rm{p}} = 15.2 \pm 5.9 M_{\oplus}$ and projected separation is 
$r_{\perp} = 1.21 \pm 0.27$ AU.
These imply that the lens system consists of a low mass star and a cold sub-Neptune.

We also estimated the probability distributions of physical parameters of the lens system 
using a Bayesian analysis with $t_{\rm{E}}$ and $\theta_{\rm{E}}$, 
which were derived from the constrained xallarap model. 
The Bayesian analysis yields that the host star is K-dwarf with mass 
of $M_{\rm{h}} = 0.64_{-0.34}^{+0.22} M_\sun$ at $D_{\rm{L}} = 4.6_{-1.8}^{+1.1}$ kpc 
and the planet mass is $M_{\rm{p}} = 109_{-58}^{+38} M_\oplus = 1.15_{-0.61}^{+0.40} M_{\rm{Saturn}}$ 
and projected separation is $r_\perp = 3.8_{-1.5}^{+0.9}$ AU. 

As mentioned in Section \ref{sec:xallarap},
the unconstrained xallarap model shows smaller $\chi^2$ value than that of the parallax plus orbital motion model, but
only by $\Delta\chi^2 = 5$ for two more degrees of freedom.
While formally significant, this difference could also be caused by rather modest 
systematic errors, and we found the parallax is preferred rather than the xallarap.
High angular resolution follow-up observations by the Hubble Space Telescope 
(HST) or Adaptive Optics (AO) can be used to confirm if the 
parallax plus orbital motion model is correct observationally.
If the AO or HST observations are conducted, they should resolve stars
unrelated to the source and lens stars that are blended with the lens
and source stars in seeing-limited ground-based images.
This should allow the brightness of the combined lens and source stars to be determined.
Because we know the source brightness from the models,
we can get the brightness of the lens by subtracting the brightness of the source
from the brightness measured by the HST or AO observations.
If the lens brightnesses derived from the parallax plus orbital motion model differ vastly from those 
derived from the xallarap model, we can confirm if the parallax plus orbital motion model is correct.
As shown in Section \ref{sec:lens}, the probability distributions of lens brightnesses of xallarap model have two peaks.
The brighter one consists of blue turn-off dwarf stars with about solar mass at the far side of the disk.
The fainter one consists of late M dwarfs with $M \sim 0.2$--$0.4 M_{\sun}$ in the  closer disk at 2--4 kpc.
On the other hand, the lens of the parallax plus orbital motion model is redder and closer than the stars in this fainter peak.
So these models can be distinguished by the brightness measurements with multiple band.
Furthermore, if the observations with HST would be conducted
after a few years, when the lens and source have separated
far enough for their relative positions to be determined \citep{bennett07}, 
we can know the direction of the lens-source motion.
Then we can confirm if the parallax plus orbital motion model is correct
by comparing the observed direction of the lens-source motion to
the parallax plus orbital motion prediction.

OGLE has started their EWS and issued 1744 microlens alerts in 2012.
Additionally, Wise Observatory in Israel began the survey observation 
with 1 m telescope equipped with a 1 deg$^{2}$ FOV camera.
The Korean Microlensing Telescope Network (KMTNet),
which is a network using three 1.6 m telescopes with 4.0 deg$^{2}$ CCD cameras, 
will provide continuous coverage of microlensing events.
These enable us to get well covered data for most microlensing events and to find more planetary events.
If we could observe the planetary anomalies without the need for follow-up observations,
the statistical analysis of the exoplanet distribution by microlensing would become easier,
because we would not need to consider the effect of the follow-up observations on the detection efficiency.
With more robust statistics we could approach understanding of explonets
from the various directions, such as the dependence of the mass of the host star.

\acknowledgments 
We acknowledge the following sources of support: The MOA project is supported by
the Grant-in-Aid for Scientific Research (JSP19340058, JSPS20340052, JSPS22403003) and
the Global COE Program of Nagoya University
``Quest for Fundamental Principles in the Universe'' from
JSPS and MEXT of Japan.
The OGLE project has received funding from the European Research Council
under the European Community's Seventh Framework Programme
(FP7/2007-2013)/ERC grant agreement no. 246678 to AU.
A. Gould acknowledges support from NSF
AST-1103471. B.S. Gaudi, A. Gould, L.-W. Hung, and R.W. Pogge
acknowledge support from NASA grant NNX12AB99G. Work by J.C. Yee was
supported by
a National Science Foundation Graduate Research Fellowship under Grant
No. 2009068160. B. Shappee and J. van Saders are also supported by
National Science Foundation Graduate Research Fellowships. Work by C. Han
was supported by Creative Research
Initiative Program (2009-0081561) of National Research Foundation of
Korea. Work by S. Dong was performed under contract with the
California Institute of Technology (Caltech) funded by NASA through
the Sagan Fellowship Program.
This work is based in part on data collected by MiNDSTEp with the Danish 
1.54m telescope at the ESO La Silla Observatory. The Danish 1.54m 
telescope is operated based on a grant from the Danish Natural Science 
Foundation (FNU). The MiNDSTEp monitoring campaign is powered by ARTEMiS 
(Automated Terrestrial Exoplanet Microlensing Search; Dominik et al. 2008, 
AN 329, 248). MH acknowledges support by the German Research Foundation 
(DFG). DR (boursier FRIA) and JSurdej acknowledge support from the 
Communaut\'{e} fran\c{c}aise de Belgique -- Actions de recherche 
concert\'{e}es -- Acad\'{e}mie universitaire Wallonie-Europe. KA, DMB, MD, 
KH, MH, CL, CS, RAS, and YT are thankful to Qatar National Research Fund 
(QNRF), member of Qatar Foundation, for support by grant NPRP 
09-476-1-078.
TCH gratefully acknowledges financial support from the Korea Research
Council for Fundamental Science and Technology (KRCF) through the
Young Research Scientist Fellowship Program. TCH acknowledges
financial support from KASI (Korea Astronomy and Space Science
Institute) grant number 2013-9-400-00.
CUL acknowledges financial support from KASI grant number 2012-1-410-02.
KH is supported by a Royal Society Leverhulme Trust Senior Research Fellowship.

\newpage

\begin{deluxetable}{crr}
\tablecolumns{3}
\tablewidth{0pc}
\tablecaption{The dataset used for the modeling}
\tablehead{
\colhead{Observatory} & \colhead{Filter} & \colhead{$N_{\rm{data}}$}}
\startdata
    MOA           &    MOA-Red &  3654 \\
    OGLE          &          I &  1436 \\
    CTIO          &          I &   118 \\
                  &          V &    10 \\
    Palomar       &          I &    26 \\
    Farm Cove     & Unfiltered &    44 \\
    SAAO          &          I &   133 \\
                  &          V &     6 \\
    Canopus       &          I &    37 \\
    Faulkes North &          I &    62 \\
                  &          V &     4 \\
    Liverpool     &          I &    37 \\
    Danish        &          I &   131 \\
\enddata
\label{tab:dataset}
\end{deluxetable}

\begin{deluxetable}{crrrrrrr}
\tabletypesize{\scriptsize}
\tablecolumns{8}
\tablewidth{0pc}
\tablecaption{Model parameters}
\tablehead{
\colhead{parameters} & \colhead{standard} & \colhead{parallax} & \colhead{unconstrained} & 
\colhead{constrained} & \colhead{orbital} & \colhead{parallax} & \colhead{parallax} \\
\colhead{} & \colhead{} & \colhead{} & \colhead{xallarap} & 
\colhead{xallarap} & \colhead{} & \colhead{+ orbital} & \colhead{+ orbital} \\
\colhead{} & \colhead{} & \colhead{} & \colhead{} & 
\colhead{} & \colhead{} & \colhead{($u_0<0$)} & \colhead{($u_0>0$)}
} 
\startdata
$t_{0}$                & 5378.641 & 5378.717 & 5378.723 & 5378.706 & 5378.776 & 5378.683 & 5378.694 \\
 (HJD')                &    0.015 &    0.017 &    0.015 &    0.013 &    0.036 &    0.014 &    0.017 \\
\cline{1-8}
$t_{\rm{E}}$           &     57.2 &     70.3 &     62.9 &     61.8 &     75.1 &     62.6 &     64.2 \\
 (day)                 &      0.3 &      0.7 &      0.3 &      0.3 &      0.9 &      0.6 &      0.6 \\
\cline{1-8}
$u_{0}$                &   0.0816 &   0.0644 &  -0.0722 &  -0.0741 &   0.0596 &  -0.0721 &   0.0716 \\
                       &   0.0005 &   0.0007 &   0.0005 &   0.0004 &   0.0007 &   0.0008 &   0.0007 \\
\cline{1-8}
$q \times 10^{4}$      &     8.16 &     4.46 &     5.17 &     5.16 &    11.63 &     2.60 &     3.68 \\
                       &     0.11 &     0.07 &     0.08 &     0.06 &     0.92 &     0.53 &     1.26 \\
\cline{1-8}
$s$                    &    1.243 &    1.192 &    1.216 &    1.220 &    1.310 &    1.154 &    1.180 \\
                       &    0.001 &    0.002 &    0.001 &    0.001 &    0.012 &    0.016 &    0.028 \\
\cline{1-8}
$\alpha$               &   0.1694 &   0.1976 &  -0.1740 &  -0.2024 &   0.1385 &  -0.2743 &   0.1965 \\
 (rad)                 &   0.0005 &   0.0010 &   0.0005 &   0.0004 &   0.0081 &   0.0087 &   0.0151 \\
\cline{1-8}
$\rho \times 10^{3}$   &     1.91 &     1.09 &     1.31 &     1.35 &     1.66 &     0.93 &     1.09 \\
                       &     0.02 &     0.02 &     0.02 &     0.01 &     0.06 &     0.10 &     0.17 \\
\cline{1-8}
$\pi_{\rm{E,N}}$       &  \nodata &     0.35 &  \nodata &  \nodata &  \nodata &     1.01 &     0.72 \\
                       &          &     0.01 &          &          &          &     0.06 &     0.05 \\
\cline{1-8}
$\pi_{\rm{E,E}}$       &  \nodata &    -0.13 &  \nodata &  \nodata &  \nodata &    -0.51 &    -0.39 \\
                       &          &     0.03 &          &          &          &     0.04 &     0.03 \\
\cline{1-8}
$\xi_{\rm{E,N}}$       &  \nodata &  \nodata &    -2.58 &     0.02 &  \nodata &  \nodata &  \nodata \\
                       &          &          &  \nodata &  \nodata &          &          &          \\
\cline{1-8}
$\xi_{\rm{E,E}}$       &  \nodata &  \nodata &    -1.86 &     0.04 &  \nodata &  \nodata &  \nodata \\
                       &          &          &  \nodata &  \nodata &          &          &          \\
\cline{1-8}
$\rm{RA}_{\xi}$        &  \nodata &  \nodata &   256.07 &   255.77 &  \nodata &  \nodata &  \nodata \\
 (deg)                 &          &          &  \nodata &  \nodata &          &          &          \\
\cline{1-8}
$\rm{decl}_{\xi}$      &  \nodata &  \nodata &   -23.44 &    -0.89 &  \nodata &  \nodata &  \nodata \\
 (deg)                 &          &          &  \nodata &  \nodata &          &          &          \\
\cline{1-8}
$P_{\xi}$              &  \nodata &  \nodata &   475.53 &   155.66 &  \nodata &  \nodata &  \nodata \\
 (day)                 &          &          &  \nodata &  \nodata &          &          &          \\
\cline{1-8}
$\epsilon$             &  \nodata &  \nodata &     0.17 &     0.20 &  \nodata &  \nodata &  \nodata \\
                       &          &          &  \nodata &  \nodata &          &          &          \\
\cline{1-8}
$\omega \times 10^{3}$ &  \nodata &  \nodata &  \nodata &  \nodata &    -0.93 &    -7.39 &    -1.39 \\
 (rad day$^{-1}$)      &          &          &          &          &     0.26 &     0.39 &     0.60 \\
\cline{1-8}
$ds/dt \times 10^{3}$  &  \nodata &  \nodata &  \nodata &  \nodata &    -5.67 &     2.51 &     1.41 \\
 (day$^{-1}$)          &          &          &          &          &     0.56 &     0.63 &     1.16 \\
\cline{1-8}
$\chi^{2}$             &  6037.32 &  5684.47 &  5651.69 &  5652.59 &  5716.16 &  5657.75 &  5660.31 \\
$dof$             &  5664 &  5662 &  5658 &  5658 &  5662 &  5660 &  5660 \\
\enddata
\tablecomments{To estimate the errors, the xallarap parameters are fixed at the best values 
because xallarap parameters are strongly degenerate. 
We assumed $M_{\rm{S}} = M_{\rm{c}} = 1 M_{\sun}$ for the constraint in the xallarap model.
}
\label{tab:parameter}
\end{deluxetable}



\end{document}